\documentclass[%
 reprint,
superscriptaddress,
 amsmath,amssymb,
 aps,
onecolumn,
showkeys,
]{revtex4-2}

\usepackage{graphicx}
\usepackage{dcolumn}
\usepackage{bm}
\usepackage{hyperref}
\usepackage{braket}
\usepackage{multirow}
\usepackage{float}
\usepackage{epstopdf}

\begin{document}


\title{Spectra and Decay Properties of Higher Lying $B_C$ Meson States}

\author{Nayana T S}
 \altaffiliation[Presently at: ]{Centre for High Energy Physics, Indian Institute of Science, Bangalore 560012, India.}
\author{Bhaghyesh}
 \email{bhaghyesh.mit@manipal.edu}
 \altaffiliation[ORCID iD: ]{0000-0003-3994-9945}
\affiliation{%
 Department of Physics, Manipal Institute of Technology,\\
 Manipal Academy of Higher Education, Manipal 576104, India.
}%





\begin{abstract}
In this work, the spectra and decay properties of $B_{c}$ mesons $( c\overline{b} )$ have been investigated using a non-relativistic potential model incorporating corrections from LQCD. The non-relativistic Schrodinger wave equation is solved numerically using the Matrix Numerov Method. Using the obtained masses and wave functions, decay widths, lifetime, branching ratios and radiative decay widths are computed for the $c\overline{b}$ system. We compare the obtained results with the experimental data and with other theoretical models.
\end{abstract}

\maketitle

\section{Introduction}
$B_{c}$ mesons are open flavored mesons consisting of a heavy quark and a heavy antiquark of dissimilar flavors ($c\bar{b}$ or $b\bar{c}$) with a net non-zero charge. Energy levels of $B_c$ mesons lies in between that of charmonium ($c\bar{c}$) and bottomonium ($b\bar{b}$). Due to similarity with hidden flavor quarkonia, $B_c$ mesons provide an opportunity to validate the potentials and formalisms used to study charmonium and bottomonim states \cite{gershtein1995physics,ortega2020spectroscopy}. Spectroscopic study of $B_c$ mesons will provide us important information on heavy quark dynamics and help us extract various parameters of the electroweak theory \cite{gershtein1995physics}. Experimentally, only very few $c\bar{b}$ states have been discovered. It is expected that more states would be discovered in near future at the B factories, the Tevatron and LHCb \cite{donati2006prospects}. Presently only three $B_{c}$ states have been experimentally observed out of which masses have been measured only for two low lying states. $B_c (1S)$ state was discovered in 1998 by the CDF Collaboration in $p\bar{p}$ collisions at the Fermilab Tevatron and estimated its mass to be $6.40\pm 0.39\pm 0.13$ GeV$/c^2$ and lifetime to be $0.46^{+0.18}_{-0.16}\pm 0.03$ ps \cite{disc1}. Signals consistent with $B_c^{+} (2S)$ and $B_c^{*+} (2S)$ were observed in proton-proton collisions by the CMS Collaboration during the 2015–2018 run at LHC \cite{disc2}. The mass of the $B_c^{+} (2S)$ state was measured to be $6871.0\pm1.2\pm 0.8\pm 0.8$ MeV \cite{disc2}. The mass of $B_c^{*+} (2S)$ was not measured in this experiment. The vector ground state $B_c^{*+} (1S)$ has not yet been detected, but the mass difference $B_c^{*+} (2S)-B_c^{*+} (1S)$ is reported to be 567 MeV \cite{disc2}. The present PDG masses of $B_c^{+} (1S)$ and $B_c^{+} (2S)$ states respectively are $6274.47\pm 0.27\pm 0.17$ MeV and $6871.2\pm 1.0$ MeV and the mean lifetime of $ B_c^{+} (1S)$ is estimated to be $0.510\pm 0.009$ ps \cite{pdg}. 

 The $c\bar{b}$ state cannot annihilate into gluons and are more stable since the flavor quantum numbers are non-vanishing, and these numbers are preserved in electromagnetic and strong interactions. Also, the pseudo-scalar ground state of $B_{c}$ meson having energies less than the BD meson production threshold can decay through weak interactions alone, making it a suitable area to examine weak decays \cite{PhysRevD.41.2856, PhysRevD.49.3399,ElHady1999bethesalpeter}. Studying the $B_{c}$ meson weak decays helps in determining the CKM elements and other Standard Model parameters from the experimental view point \cite{PhysRevD.49.3399,Choi_1998}. Excited $B_{c}$ states below BD threshold can only undergo hadronic or radiactive decay to the ground state which will further decay through weak interactions. These are considerably stable as compared to the corresponding $c\overline{c}$ and $b\overline{b}$ states \cite{eichten1994mesons}. If the mesons possess an energy above the BD threshold, then it is most likely that they will decay into lighter mesons. Higher stability of $c\bar{b}$ state is the cause for smaller decay widths (few hundred KeV) of $B_{c}$ mesons. 

\noindent Theoretically, there are various approaches employed to study $c\overline{b}$ systems in literature which include QCD sum rules \cite{bcspectroscopy}, heavy quark effective theory \cite{zeng1995heavy}, lattice QCD \cite{hpqcd2005mass}, Bethe-Salpeter equation \cite{ElHady1999bethesalpeter,PhysRevD.100.054034}, phenomenological potential models \cite{eichten1994mesons,Ebert:2011jc}, etc. The predictions from these approaches can be validated once further experimental data on these mesons are obtained. In the present article, we have computed the spectra and decay properties of $B_c$ mesons using a non relativistic potential model. Phenomenological potential model approach is one of the important methods used to investigate heavy meson systems and is very successful \cite{PhysRevLett.34.369, PhysRevD.17.3090,PhysRevD.70.054017, PhysRevD.74.014012, PhysRevD.26.3305, PhysRevD.75.074031, PhysRevD.79.094004, barnes2005higher,chaturvedi2018massspectra}. Most of the potential models involve solving the non relativistic Schrodinger equation for the chosen $q\bar{q}$ potential. Since these solutions often cannot be obtained analytically, we need to employ numerical methods. Various numerical methods used in potential models are: approximations based on Runge-Kutta method \cite{chaturvedi2018massspectra}, matrix Numerov method \cite{manzoor2021newvariational}, discrete variable representation (DVR) method \cite{bhagAHEP}, Fourier grid Hamiltonian method \cite{brau1998three}, asymptotic iteration method \cite{mutuk2019spin}, variational method \cite{rai2006properties}, artificial neural network method \cite{mutuk2019cornell}, etc. Among these, the Numerov method is a simple but efficient method to solve Schrodinger equation \cite{pillai2012matrix}. In this method both the kinetic energy and potential energy are represented as simple matrices on a discretized lattice. Then the problem of solving the Schrodinger equation reduces to solving a matrix eigenvalue problem. The matrix Numerov method have been successfully applied to heavy mesons like charmonium and bottomonium \cite{manzoor2021newvariational}. However, this method has not been applied to open flavored mesons. In this article, we numerically solve the Schrödinger equation for $B_c$ mesons using the matrix Numerov method.

\noindent This article is arranged as follows: in Section \ref{sec 2} we present the potential model used in our analysis and we also briefly discuss the matrix Numerov method used to solve the Schrodinger equation. The decay properties computed in the present work are given in Section \ref{sec 3}. Results and discussions of the present work are given in Section \ref{sec 4}.

\section{Potential Model}\label{sec 2}

In this work, we have employed a non-relativistic model to study the properties of $c\overline{b}$ system. Our aim is to solve the nonrelativistic Schrodinger wave equation for the two particle $c\bar{b}$ system:
\begin{equation}
H\psi = E\psi \label{swe}
\end{equation}
The model Hamiltonian takes the form
\begin{equation}
H = M + \frac{p^2}{2\mu} + V(r) ,\label{hamil}
\end{equation}
\noindent where $M = m_c+m_{\bar{b}}$, $p$ is the relative momentum, $\mu~(= m_c m_{\bar{b}}/m_c+m_{\bar{b}})$ is the reduced mass of the $c\bar{b}$ system, where $m_c$ and $m_{\bar{b}}$ are the masses of $c$ quark and $\bar{b}$ antiquark respectively. In Eq.\ref{hamil}, the quark-antiquark potential $V(r)$ is taken to be of the following form:
\begin{equation}
    V(r) = -\frac{4\alpha_{s}}{3r} + Ar + B\ln(r\lambda)  + \frac{32\pi\alpha_{s}}{9m_{c}m_{\overline{b}}}\left(\frac{\sigma}{\sqrt{\pi}}\right)^{3}e^{-\sigma^{2}r^{2}}\mathbf{S_{1}}.\mathbf{S_{2}}\label{pot}
\end{equation}
The first two terms together in Eq.\ref{pot} is called the Cornell potential \cite{PhysRevLett.34.369, PhysRevD.17.3090, barnes2005higher} and has been successfully used in quarkonium spectroscopy. The first term of the Cornell potential (generally called the Coulombic term) represents the short distance behaviour and can be derived from the one-gluon-exchange interaction in QCD. The linearly increasing second term models confinement, which is a nonperturbative effect prominent at large inter-quark separation. Even though the specific form of quark-antiquark potential for all distances has not been derived from QCD, various studies \cite{sumino1, sumino2, Andreev, white, kinar}, including Lattice QCD in the static limit \cite{bali, lattice2}, indicate the central spin independent QCD potential to be of Cornell type. Using lattice QCD, various authors \cite{koma, koma2, kawanai, laschka, laschka2} have derived the relativistic corrections to the static potential at order $1/m$ in the quark mass. In these studies the correction term was found to be proportional to $\ln(r)$, where $r$ is the inter-quark separation. Such logarithmic correction term has also been suggested using effective string theory \cite{nadal}. This $\mathcal{O}(1/m)$ correction term has been proved to influence charmonium and bottomonium spectra \cite{kawanai, laschka}. Hence, we include this term in our model potential. The last term in Eq.\ref{pot} represents a Gaussian-smeared contact hyperfine interaction \cite{barnes2005higher}. The spin-hyperfine interaction is treated non-perturbatively in the present analysis. The values of parameters used in our model are given in Table \ref{tab1}.

\begin{table}[h]
\caption{List of parameters.\label{tab1}}
\begin{ruledtabular}
\begin{tabular}{cccc}
Parameter     & Value     & Parameter     & Value     \\
\colrule
$m_c$      & 1.46 GeV      & $m_{\bar{b}}$      & 4.68 GeV      \\ 
$A$      & 0.8126 GeV$^2$      & $B$      & 0.0913 GeV      \\
$\lambda$      & 0.1210 GeV      & $\sigma$      & 3.7374 GeV      \\
\end{tabular}
\end{ruledtabular}
\end{table}

\noindent The QCD coupling constant $\alpha_s$ in Eq.\ref{pot} is computed using the formula 
\begin{equation*}
\alpha_{s}\left(\mu_{0}^{2}\right) = \frac{4\pi}{\left(11 - \frac{2n_{f}}{3}\right)\ln\left(\frac{\mu_{0}^{2} + M_{B}^{2}}{\Lambda^{2}}\right)},  \end{equation*}
where $\Lambda$ ($=0.168$ GeV) is the characteristic scale of QCD \cite{ebert2003}, $\mu_{0}=2\mu$ with $\mu$ being the reduced mass, $n_{f}~(=4)$ is the number of flavors of quarks having mass lower than $b$ quark mass, $M_{B}~(=1 $ GeV) is the background mass \cite{badalian2001, badalian2006}.
The energies and wavefunctions of $c\bar{b}$ system can be obtained by solving the Schrödinger equation (Eq.\ref{swe}) using the potential given in Eq.\ref{pot}. 

\begin{figure}[h]
\centering
\includegraphics[scale=0.75]{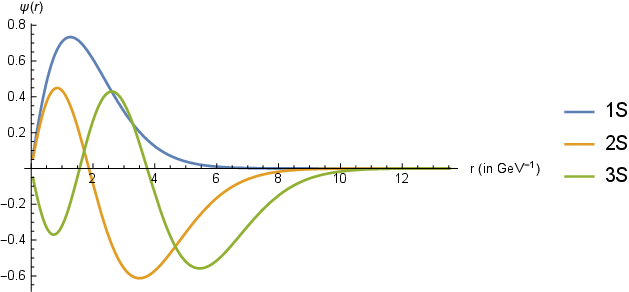}
\includegraphics[scale=0.75]{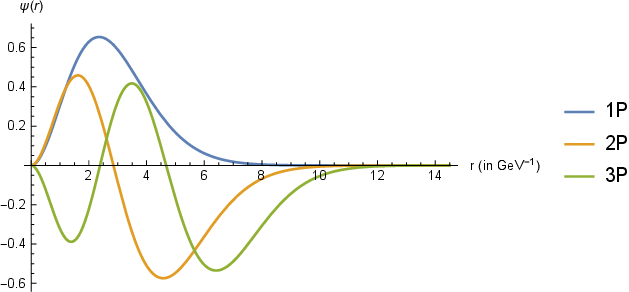}
\includegraphics[scale=0.75]{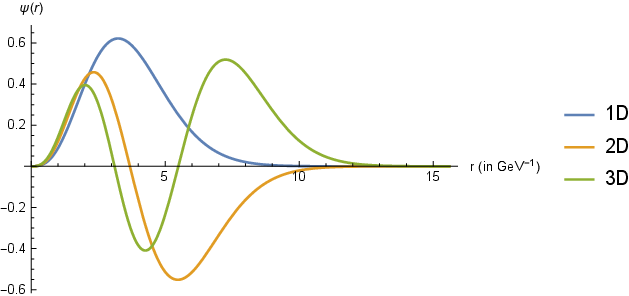}
\includegraphics[scale=0.75]{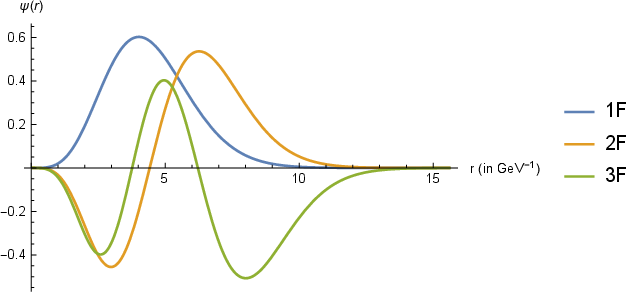}
\caption{Wave-functions of first three S-states, P-states, D-states and F-states of $B_{c}$ meson}
\label{plot1}
\end{figure}

\noindent In the present article, we have used the matrix Numerov method to numerically solve the Schrodinger equation. In this method \cite{pillai2012matrix}, the radial coordinate $r$ is discretized into $N$ equidistant lattice points $r_i$. The kinetic energy operator appearing in the Schrodinger equation is represented by two trigonal matrices and the potential energy operator is a diagonal matrix whose diagonal entries are the values of potential energy evaluated at each lattice point. Thus, the radial Schrodinger equation takes the form of a matrix eigenvalue equation given by \cite{manzoor2021newvariational, pillai2012matrix}
\begin{equation}
-\frac{1}{2\mu}A_{N,N}B_{N,N}^{-1} \psi_{i} + [V_{N}(r_{i}) + \frac{l(l+1)}{2\mu r_{i}^{2}} + m_{c} + m_{b}]\psi_{i} = E\psi_{i}. \label{mnm}
\end{equation}
In Eq.\ref{mnm}, $A$ and $B$ are ($N\times N$) trigonal matrices representing the kinetic energy operator and are given by \cite{manzoor2021newvariational, pillai2012matrix}
\begin{eqnarray}
A_{N,N} &=& \frac{I_{-1} - 2I_{0} + I_{1}}{d^{2}}  \nonumber\\
B_{N,N} &=& \frac{I_{-1} + 10I_{0} + I_{1}}{12},
\end{eqnarray}
where, $I_{0}, I_{-1}, I_{1}$ are the identity matrix, lower shift identity matrix and upper shift identity matrix respectively, all of order N $\times$ N and $d$ is the grid spacing. $V_N (=V(r_i)\delta_{ij})$ is the $N\times N$ diagonal potential energy matrix and $\psi_i = \psi(r_i)$ is an $N$ dimensional column matrix. The eigenvalue problem corresponding to Eq.\ref{mnm} was solved using Mathematica taking $N=350$ and we obtain the bound state eigenvalues ($E$) and values of the corresponding wavefunctions ($\psi_i$) at the chosen grid points. Computed wavefunctions for $S$, $P$, $D$ and $F$ states are shown in Figure \ref{plot1}. 

\noindent For computing the fine structure of the $L \neq 0$ states, the spin-orbit and tensor terms are added perturbatively to the obtained eigenvalues. The spin-orbit term chosen is of the form \cite{gershtein1995physics,PhysRevD.70.054017,yangmao}
\begin{equation*}
V_{LS}(r) = \frac{4\alpha_{s}}{3r^3}\left(\frac{1}{m_c}+\frac{1}{m_b}\right) \left(\frac{S_c\cdot L}{m_c}+\frac{S_b\cdot L}{m_b}\right)
-\frac{1}{2r}\frac{\partial V^{conf}(r)}{\partial r}\left(\frac{S_c\cdot L}{m_c^2}+\frac{S_b\cdot L}{m_b^2}\right) .
    \end{equation*}
The tensor term used is of the form \cite{gershtein1995physics,PhysRevD.70.054017,yangmao}\\
\begin{equation*}
V_{T} = \frac{4}{3}\frac{\alpha_s}{m_c m_b}\frac{1}{r^3}\left( 3S_c\cdot \hat{r}~S_b\cdot \hat{r} - S_c\cdot S_b  \right) .
\end{equation*} 
For mesons with unequal quark masses, the spin-orbit interaction $V_{LS}$ in the Hamiltonian causes mixing of the $^1L_J$ and $^3L_J $ states that have the same $J^P$ quantum numbers. The physical states are linear combination of these two states. The spin-dependent corrections are evaluated perturbatively in the $\ket{JM,LS}$ basis \cite{PhysRevD.70.054017,yangmao}. The mass matrix is constructed in this basis and then diagonalized to obtain the mixing eigenstates \cite{yangmao}.  The spectra of $B_c$ mesons obtained from our analysis are listed in Tables \ref{tab_sp} \& \ref{tab_df}. The mixing eigenstates are listed in Table \ref{mix}. 
We have also computed the root mean square radii $\sqrt{\langle r^{2} \rangle_{nl}} = \sqrt{\int_{0}^{\infty}r^{4}|R_{nl}(r)|^{2}}$ and the results are listed in Table \ref{tab_rms}.

\begin{table}[h]
\caption{Spectra of $S$ and $P$ states (in GeV).\label{tab_sp}}
\begin{ruledtabular}
\begin{tabular}{cccccccccc}
SState& Ours & Exp.\cite{pdg} & \cite{Lat} & \cite{PhysRevD.70.054017} & \cite{Ebert:2011jc} & \cite{ABREU} & \cite{QLI} & \cite{soni} & \cite{XJLi} \\
\colrule
$B_c(1^1S_0)$& 6.273 & 6.27447$\pm$0.00027  & 6.276(3)(6) & 6.271 & 6.272 & 6.277 & 6.271 & 6.272 & 6.271 \\
$ B_c(2^1S_0)$& 6.879 & 6.871$\pm$0.001  &   & 6.855 & 6.842 & 6.845 & 6.871 & 6.864 & 6.855 \\
$ B_c(3^1S_0)$& 7.303 &   &   & 7.250 & 7.226 & 7.284 & 7.239 & 7.306 & 7.220 \\
$ B_c(4^1S_0)$& 7.658 &   &   &   & 7.585 &   & 7.540 & 7.684 & 7.496 \\
$ B_c(5^1S_0)$& 7.973 &   &   &   & 7.928 &   & 7.805 & 8.025 & 7.722 \\
$ B_c(6^1S_0)$& 8.262 &   &   &   &   &   & 8.046 & 8.340 &   \\
$ B_c(1^3S_1)$& 6.328 &   & 6.331(4)(6) & 6.338 & 6.333 & 6.288 & 6.326 & 6.321 & 6.338 \\
$ B_c(2^3S_1)$& 6.912 &   &   & 6.887 & 6.882 & 6.853 & 6.890 & 6.900 & 6.886 \\
$ B_c(3^3S_1)$& 7.330 &   &   & 7.272 & 7.258 & 7.290 & 7.252 & 7.338 & 7.240 \\
$ B_c(4^3S_1)$& 7.681 &   &   &   & 7.609 &   & 7.550 & 7.714 & 7.512 \\
$ B_c(5^3S_1)$& 7.995 &   &   &   & 7.947 &   & 7.813 & 8.054 & 7.735 \\
$ B_c(6^3S_1)$& 8.282 &   &   &   &   &   & 8.054 & 8.368 &   \\
$ B_c(1^3P_0)$& 6.677 &   & 6.712(18)(7) & 6.706 & 6.699 & 6.639 & 6.714 & 6.686 & 6.701 \\
$ B_c(1^3P_2)$& 6.751 &   &   & 6.768 & 6.761 & 6.667 & 6.787 & 6.712 & 6.773 \\
$ B_c(1P^\prime)$& 6.744 &   &   & 6.750 & 6.750 & 6.656 & 6.776 & 6.705 & 6.754 \\
$ B_c(1P)$& 6.729 &   & 6.736(17)(7) & 6.741 & 6.743 & 6.606 & 6.757 & 6.706 & 6.745 \\
$ B_c(2^3P_0)$& 7.122 &   &   & 7.122 & 7.094 & 7.123 & 7.107 & 7.146 & 7.097 \\
$ B_c(2^3P_2)$& 7.192 &   &   & 7.164 & 7.157 & 7.127 & 7.160 & 7.173 & 7.148 \\
$ B_c(2P^\prime)$& 7.186 &   &   & 7.150 & 7.147 & 7.121 & 7.150 & 7.165 & 7.133 \\
$ B_c(2P)$& 7.168 &   &   & 7.145 & 7.134 & 7.088 & 7.134 & 7.168 & 7.125 \\
$ B_c(3^3P_0)$& 7.489 &   &   &   & 7.474 & 7.523 & 7.420 & 7.536 & 7.393 \\
$ B_c(3^3P_2)$& 7.558 &   &   &   & 7.524 & 7.515 & 7.464 & 7.565 & 7.434 \\
$ B_c(3P^\prime)$& 7.551 &   &   &   & 7.510 & 7.513 & 7.458 & 7.555 & 7.421 \\
$ B_c(3P)$& 7.533 &   &   &   & 7.500 & 7.488 & 7.441 & 7.559 & 7.414 \\
$ B_c(4^3P_0)$& 7.813 &   &   &   & 7.817 &   & 7.693 & 7.885 & 7.633 \\
$ B_c(4^3P_2)$& 7.881 &   &   &   & 7.867 &   & 7.732 & 7.915 & 7.667 \\
$ B_c(4P^\prime)$& 7.874 &   &   &   & 7.853 &   & 7.727 & 7.905 & 7.656 \\
$ B_c(4P)$& 7.855 &   &   &   & 7.844 &   & 7.710 & 7.908 & 7.650 \\
\end{tabular}
\end{ruledtabular}
\end{table}

\begin{table}[h!]
\caption{Spectra of $D$ and $F$ states (in GeV).\label{tab_df}}
\begin{ruledtabular}
\begin{tabular}{ccccccccc}
State& Ours & Exp.\cite{pdg} & \cite{PhysRevD.70.054017} & \cite{Ebert:2011jc} & \cite{ABREU} & \cite{QLI} & \cite{soni} & \cite{XJLi} \\
\colrule
$B_c(1^3D_1)$& 7.019 &   & 7.028 & 7.021 &   & 7.020 & 6.998 & 7.023 \\
$B_c(1^3D_3)$& 7.021 &   & 7.045 & 7.029 &   & 7.030 & 6.990 & 7.042 \\
$B_c(1D^\prime)$& 7.031 &   & 7.036 & 7.026 & 6.920 & 7.032 & 6.997 & 7.039 \\
$B_c(1D)$& 7.018 &   & 7.028 & 7.025 & 6.931 & 7.024 & 6.994 & 7.032 \\
$B_c(2^3D_1)$& 7.398 &   &   & 7.392 &   & 7.336 & 7.403 & 7.327 \\
$B_c(2^3D_3)$& 7.404 &   &   & 7.405 &   & 7.348 & 7.399 & 7.344 \\
$B_c(2D^\prime)$& 7.409 &   &   & 7.400 & 7.345 & 7.347 & 7.403 & 7.340 \\
$B_c(2D)$& 7.401 &   &   & 7.399 & 7.334 & 7.343 & 7.401 & 7.335 \\
$B_c(3^3D_1)$& 7.730 &   &   & 7.732 &   & 7.611 & 7.762 & 7.573 \\
$B_c(3^3D_3)$& 7.739 &   &   & 7.750 &   & 7.625 & 7.761 & 7.589 \\
$B_c(3D^\prime)$& 7.741 &   &   & 7.743 & 7.704 & 7.623 & 7.764 & 7.584 \\
$B_c(3D)$& 7.736 &   &   & 7.741 & 7.694 & 7.620 & 7.762 & 7.581 \\
$B_c(1^3F_2)$& 7.269 &   & 7.269 & 7.273 &   & 7.235 & 7.234 & 7.252 \\
$B_c(1^3F_4)$& 7.251 &   & 7.271 & 7.277 &   & 7.227 & 7.244 & 7.253 \\
$B_c(1F^\prime)$& 7.273 &   & 7.266 & 7.269 &   & 7.240 & 7.242 & 7.260 \\
$B_c(1F)$& 7.251 &   & 7.276 & 7.268 &   & 7.224 & 7.241 & 7.248 \\
$B_c(2^3F_2)$& 7.611 &   &   & 7.618 &   & 7.518 & 7.607 & 7.507 \\
$B_c(2^3F_4)$& 7.598 &   &   & 7.617 &   & 7.514 & 7.617 & 7.510 \\
$B_c(2F^\prime)$& 7.616 &   &   & 7.616 &   & 7.525 & 7.615 & 7.514 \\
$B_c(2F)$& 7.597 &   &   & 7.615 &   & 7.508 & 7.614 & 7.505 \\
$B_c(3^3F_2)$& 7.920 &   &   &   &   & 7.730 & 7.946 &   \\
$B_c(3^3F_4)$& 7.910 &   &   &   &   & 7.771 & 7.956 &   \\
$B_c(3F^\prime)$& 7.925 &   &   &   &   & 7.779 & 7.954 &   \\
$B_c(3F)$& 7.909 &   &   &   &   & 7.768 & 7.953 &   \\
\end{tabular}
\end{ruledtabular}
\end{table}

\begin{table}[h]
\caption{{Mixing eigenstates.\label{mix}}}
\begin{ruledtabular}
\begin{tabular}{cccc}
State&\centering{Multiplet}&State&Multiplet\\
\colrule
$1P'$&$-0.728916\ket{1^1P_1}-0.684603\ket{1^3P_1}$&$2D'$&$-0.55469\ket{1^1D_2}+0.832057\ket{1^3D_2}$\\
$1P	$&$~~0.684603\ket{1^1P_1}-0.728916\ket{1^3P_1}$&$2D$&$-0.832057\ket{1^1D_2}-0.55469\ket{1^3D_2}$\\
$2P'$&$-0.746111\ket{2^1P_1}+0.665822\ket{2^3P_1}$&$3D'$&$-0.524727\ket{1^1D_2}+0.851271\ket{1^3D_2}$\\
$2P	$&$-0.665822\ket{2^1P_1}-0.746111\ket{2^3P_1}$&$3D$&$-0.851271\ket{1^1D_2}-0.524727\ket{1^3D_2}$\\
$3P'$&$-0.75317\ket{3^1P_1}	-0.657826\ket{3^3P_1}$&$1F'$&$~~0.63315\ket{1^1F_3}+0.774029\ket{1^3F_3}$\\
$3P	$&$~0.657826\ket{3^1P_1}-0.75317~~\ket{3^3P_1}$&$1F$&$-0.774029\ket{1^1F_3}+0.63315\ket{1^3F_3}$\\
$4P'$&$-0.756891\ket{4^1P_1}+0.653541\ket{4^3P_1}$&$2F'$&$-0.631225\ket{1^1F_3}+0.7756\ket{1^3F_3}$\\
$4P	$&$-0.653541\ket{4^1P_1}-0.756891\ket{4^3P_1}$&$2F$&$-0.7756\ket{1^1F_3}-0.631225\ket{1^3F_3}$\\
$1D'$&$-0.574781\ket{1^1D_2}+0.818308\ket{1^3D_2}$&$3F'$&$-0.629269\ket{1^1F_3}+0.777187\ket{1^3F_3}$\\
$1D$&$-0.818308\ket{1^1D_2}-0.574781\ket{1^3D_2}$&$3F$&$-0.777187\ket{1^1F_3}-0.629269\ket{1^3F_3}$\\
\end{tabular}
\end{ruledtabular}
\end{table}

\begin{table}[h]
\caption{RMS radii of $B_c$ meson states (in fm).\label{tab_rms}}
\begin{ruledtabular}
\begin{tabular}{cccc}
State& Ours  & \cite{Ebert:2011jc} & \cite{akbar_epja}\\
\colrule
$B_c(1^1S_0)$& 0.32 & 0.33 & 0.318 \\
$ B_c(2^1S_0)$& 0.68 & 0.63 & 0.723 \\
$ B_c(3^1S_0)$& 0.97 & 0.87 & 1.052 \\
$ B_c(4^1S_0)$& 1.22 & 1.05 & 1.337 \\
$ B_c(5^1S_0)$& 1.45 & 1.21 &   \\
$ B_c(6^1S_0)$& 1.66 &   &   \\
$ B_c(1^3S_1)$& 0.35 &   & 0.334 \\
$ B_c(2^3S_1)$& 0.70 &   & 0.732 \\
$ B_c(3^3S_1)$& 0.98 &   & 1.059 \\
$ B_c(4^3S_1)$& 1.24 &   & 1.342 \\
$ B_c(5^3S_1)$& 1.46 &   &   \\
$ B_c(6^3S_1)$& 1.68 &   &   \\
$ B_c(1P)$& 0.55 & 0.53 & 0.562 \\
$ B_c(2P)$& 0.86 & 0.79 & 0.920 \\
$ B_c(3P)$& 1.12 & 0.99 & 1.220 \\
$ B_c(4P)$& 1.36 & 1.16 &   \\
$ B_c(1D)$& 0.71 & 0.67 & 0.752 \\
$ B_c(2D)$& 1.00 & 0.90 & 1.083 \\
$ B_c(3D)$& 1.25 & 1.08 & 1.364 \\
$ B_c(1F)$& 0.86 &   &   \\
$ B_c(2F)$& 1.13 &   &   \\
$ B_c(3F)$& 1.37 &   &   \\
\end{tabular}
\end{ruledtabular}
\end{table}

\section{Decay Properties}\label{sec 3}
Using the obtained wave functions, we calculate various decay properites of $B_c$ mesons. The decays of $c\bar{b}$ states provide important information on heavy quark dynamics and allows us to extract some Standard Model parameters like the CKM elements.
\subsection{Decay Constants}
Decay constant is an important observable characterising the $B_c$ meson states \cite{gershtein1995physics}. We use the Van-Royen-Weisskopf formula \cite{van1967hardon} to calculate the decay constant \\
\begin{equation}
    f_{B_{c}} = \sqrt{\frac{3|R(0)|^{2}}{\pi M_{B_{c}}}}= \sqrt{\frac{12|\psi(0)|^{2}}{ M_{B_{c}}}}
\end{equation}
The decay constant depends on the radial wave-function at the origin, $|R(0)|^{2}$ and on the mass of corresponding $B_{c}$ meson, $M_{B_{c}}$. Including the first order QCD correction \cite{barbieri1975r,celmaster1979lepton} to the Van-Royen-Weisskopf formula, the expression for decay constant becomes\\
\begin{equation}
   \bar{f}_{B_{c}} = \sqrt{\frac{3\mid R(0)^{2}\mid}{\pi M_{B_{c}}}}\left\lbrace 1 - \frac{\alpha_{s}}{\pi}\left(\Delta - \frac{m_{c} - m_{\overline{b}}}{m_{c} + m_{\overline{b}}}\ln \mid\frac{m_{c}}{m_{\overline{b}}}\mid\right)\right\rbrace 
\end{equation}
where $\Delta$ take the values 8/3 for $^{3}S_{1}$ states and 2 for $^{1} S_{0}$ states. Table \ref{dec_con} lists $|R(0)|^{2}$, and the decay constants with and without corrections for different $c\overline{b}$ states.

\begin{table}[h]
\caption{Radial wave function at origin (in GeV$^3$) and decay constants (in GeV).\label{dec_con}}
\begin{ruledtabular}
\begin{tabular}{cccccccccc}
State& $|R(0)|^2$ & \cite{akb2} & $f_{B_{c}}$ & $\bar{f}_{B_{c}}$ & \cite{akb2} & \cite{soni} & \centering \cite{2} & \cite{3} & \cite{4}\cite{5} \\
\colrule
$B_c(1^1S_0)$& 2.835 & 2.2393 & 0.657 & 0.535 & 0.484 & 0.433 & 0.439(30)(17) & 523(62) & 528 \\
$ B_c(2^1S_0)$& 1.662 & 1.2606 & 0.480 & 0.391 & 0.347 & 0.355 & 0.282(13)(10) &   &   \\
$ B_c(3^1S_0)$& 1.357 & 1.0399 & 0.421 & 0.343 & 0.306 & 0.326 & 0.237(6)(14) &   &   \\
$ B_c(4^1S_0)$& 1.204 & 0.9358 & 0.387 & 0.316 & 0.284 & 0.307 &   &   &   \\
$ B_c(5^1S_0)$& 1.108 & 0.8726 & 0.364 & 0.297 & 0.269 & 0.294 &   &   &   \\
$ B_c(6^1S_0)$& 1.040 & 0.8293 & 0.347 & 0.283 & 0.258 & 0.284 &   &   &   \\
$ B_c(1^3S_1)$& 1.352 & 1.9566 & 0.452 & 0.368 & 0.405 & 0.435 & 0.417(51)(27) & 474(42) & 384 \\
$ B_c(2^3S_1)$& 0.898 & 1.2018 & 0.352 & 0.287 & 0.304 & 0.356 & 0.297(35)(3) &   &   \\
$ B_c(3^3S_1)$& 0.772 & 1.0083 & 0.317 & 0.258 & 0.270 & 0.326 & 0.257(29)(5) &   &   \\
$ B_c(4^3S_1)$& 0.709 & 0.9138 & 0.297 & 0.242 & 0.251 & 0.308 &   &   &   \\
$ B_c(5^3S_1)$& 0.671 & 0.8553 & 0.283 & 0.231 & 0.239 & 0.295 &   &   &   \\
$ B_c(6^3S_1)$& 0.643 & 0.8148 & 0.272 & 0.222 & 0.229 & 0.285 &   &   &   \\
\end{tabular}
\end{ruledtabular}
\end{table}

\subsection{Weak Decays}
\begin{figure*}[h]
\centering
\includegraphics[scale=0.4]{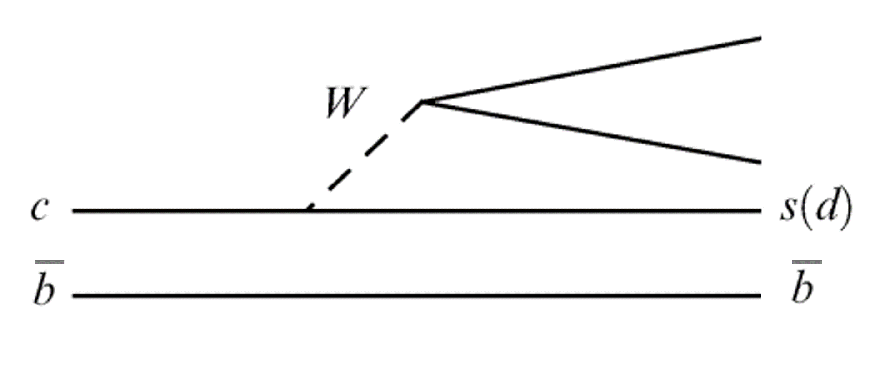}
\includegraphics[scale=0.4]{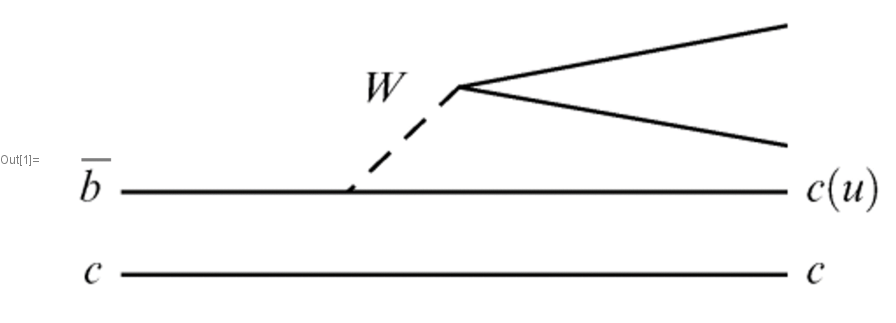}
\includegraphics[scale=0.4]{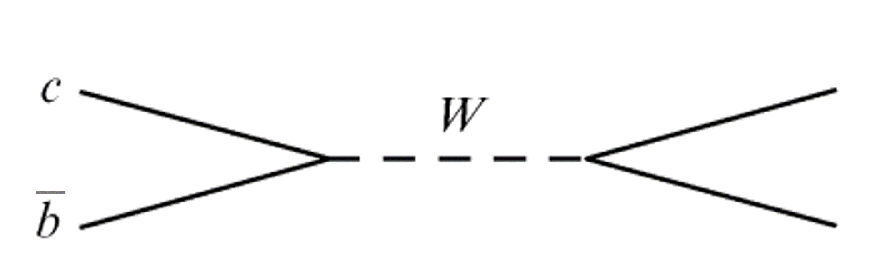}
\caption{c-quark decay, b-antiquark decay and annihilation of $B_{c}$ mesons}
\label{plot2}
\end{figure*}

We have analysed the weak decays of the low lying pseudoscalar states using a widely studied model called the spectator model \cite{gershtein1995physics, PhysRevD.49.3399, Kiselev:2003mp}. According to this model, the decay width of $B_{c}$ mesons is obtained as a result of three major processes: a) $c$-quark decay with $b$-antiquark as spectator, b) $b$-antiquark decay with $c$-quark as spectator and c) the annihilation decay of $c\overline{b}$ meson. These processes are represented \cite{gershtein1995physics} in Figure \ref{plot2}. 

During the decay, $c$-quark can decay into strange ($s$) or down ($d$) quark and simultaneously also decay into lepton-neutrino pair through the exchange of $W^{+}$ boson. Similarly $b$-antiquark decays into either charm ($c$) or up ($u$) quark and simultaneously into lepton-neutrino pair with the exchange of $W^{+}$ boson. The annihilation process of $B_{c}$ meson yields a quark-antiquark pair or a lepton-neutrino pair, ie., $B_{c}^{+} \rightarrow l^{+} \nu_{l},(c\overline{s},u\overline{s})$ where $l= e, \mu, \tau$ \cite{Abd_El_Hady_2000}. The decay width for each of these three process can be calculated using the following formulae:

\begin{eqnarray}
  \Gamma\left( c\rightarrow X \right) &=& \frac{5 G_{F}^{2}\mid V_{cs}\mid^{2}m_{c}^{5}}{192\pi^{3}}\nonumber\\
\Gamma\left( b\rightarrow X \right) &=& \frac{9 G_{F}^{2}\mid V_{cb}\mid^{2}m_{b}^{5}}{192\pi^{3}}\nonumber 
  \end{eqnarray}
\begin{equation}
 \Gamma (ann) =\sum_i\frac{G_{F}^{2}}{8\pi}\mid V_{cb}\mid^{2}f_{B_{c}}^{2}M_{B_{c}}m_{i}^{2}(1 - \frac{m_{i}^{2}}{M_{B_{c}}^{2}})C_{i}
\label{anni} 
\end{equation}

\noindent In the above equations, the values of the Cabibo-Kobayashi-Maskawa (CKM) matrix elemets $|V_{cs}|= 0.975$ and $|V_{cb}|= 0.0408$ are taken from PDG \cite{pdg}. $ G_{f} = 1.1663788 \times 10^{-5}~GeV^{-2}$ is the Fermi constant \cite{pdg}. 
%
For annihilation decay, we have considered two main channels, that is $c\overline{s}$ and $\tau \nu_{\tau}$. In Eq.\ref{anni}, $C_{i} = 3\mid V_{cs}\mid^{2}$ for $c\overline{s}$ channel and $C_{i}$ = 1 for $\tau \nu_{\tau}$ channel \cite{gershtein1995physics}. These formulae do not take into account the quark binding effects \cite{gershtein1995physics,ElHady1999bethesalpeter}. The total decay width can be written as the sum of decay widths for all the three processes \cite{gershtein1995physics}:
\begin{equation*}
   \Gamma\left( B_{c} \rightarrow X \right) = \Gamma\left( b\rightarrow X \right) + \Gamma\left( c\rightarrow X \right) + \Gamma (ann) 
\end{equation*}
The decay lifetime ($\tau$) is then given by
\begin{equation*}
   \tau = \frac{\hbar}{\Gamma}, 
\end{equation*}
where $\hbar$ = 6.582 $\times 10^{-25}$ GeV.sec. The lifetime directly depends on the decay modes.
\\
The branching ratio ($\mathcal{B}$) is the probability of decay through a particular channel. The branching ratios for $c$-decay, $\overline{b}$-decay and decay due to annihilation is calculated by dividing the decay width corresponding to a particular process by the total decay width:
\begin{eqnarray}
\mathcal{B} (c\rightarrow X) = \frac{\Gamma (c \rightarrow X)}{\Gamma(B_{c} \rightarrow X)}\nonumber\\       
\mathcal{B} (b\rightarrow X) = \frac{\Gamma (b \rightarrow X)}{\Gamma(B_{c} \rightarrow X)}\nonumber\\
\mathcal{B} (ann) = \frac{\Gamma (ann)}{\Gamma(B_{c} \rightarrow X)} \label{br}
\end{eqnarray}
The results obtained for the annihilation decay widths and total decay widths for two different channels and the corresponding lifetimes and branching ratios are listed down in Table \ref{weak}. Here, the branching ratios for Refs. \cite{ElHady1999bethesalpeter}, \cite{BHAT} and \cite{rai2006properties} have been evaluated by using their corresponding $\Gamma$ values using Eqns. \ref{br}.

\begin{table}[h]
\caption{Comparison of weak partial decay and annihilation widths (in $10^{-3}~$eV), total width (in $10^{-3}~$eV), lifetime (in ps) and branching ratios.\label{weak}}
\begin{ruledtabular}
\begin{tabular}{ccccccccc}
& $\Gamma (c\rightarrow X)$& $\Gamma (b\rightarrow X)$& $\Gamma (ann)$& $\Gamma (B_c\rightarrow X)$& $\tau$& $\mathcal{B} (c\rightarrow X)$& $\mathcal{B} (b\rightarrow X)$& $\mathcal{B} (ann)$\\
\colrule
ours & 0.72 & 0.77 & 0.13 & 1.62 & 0.41 & 0.44 & 0.47 & 0.08 \\
Exp.\cite{pdg} &   &   &   &   & 0.510$\pm$ 0.009  &   &   &   \\
\cite{ElHady1999bethesalpeter} & 0.51  & 0.75 & 0.14 & 1.40 & 0.47 & 0.36 & 0.54 & 0.10 \\
\cite{PhysRevD.70.054017} & 0.33 & 0.48 & 0.067 & 0.88 & 0.75 & 0.38 & 0.54 & 0.08 \\
\cite{BHAT} & 0.8958 & 1.0410 & 0.0057 & 1.9425 & 0.339 & 0.461 & 0.536 & 0.003 \\
\cite{rai2006properties} & 0.875 & 0.419 & 0.0923 & 1.3863 & 0.47 & 0.63 & 0.30 & 0.07 \\
\end{tabular}
\end{ruledtabular}
\end{table}

\subsection{Radiative decays}
The electric dipole (E1) and magnetic dipole (M1) radiative transitions that appears from the multipole expansions help us understand the $c\overline{b}$ system better. The $B_{c}$ mesons below the BD threshold can undergo radiative decay to the ground state which further decays through weak interactions. In E1 transitions $\Delta S = 0$ and $ |\Delta L| = 1$. The partial decay widths due to E1 transitions are calculated using \cite{PhysRevD.70.054017}
\begin{equation*}
\Gamma_{E1}(i\rightarrow f \gamma) = \frac{ 4\alpha \langle e_{Q}\rangle^{2}\omega^{3}}{3}\mid\langle f\mid r\mid i\rangle \mid^{2}C_{f_{i}}\delta_{SS'},\nonumber
\end{equation*}
where  $i$ $(\equiv n^{2S+1}L_{J})$ is the initial state and $f$ $(\equiv n'^{2S'+1}L_{J'})$ is the final state, $\langle e_{Q}\rangle = \frac{m_{\overline{b}}e_{c} - m_{c}e_{\overline{b}}}{m_{b} + m_{c}}$ is the mean charge with $e_{c} = \frac{2}{3}$ and $e_{\overline{b}} = \frac{1}{3}$ being the charges of $c$-quark and $b$-antiquark respectively. The photon energy $\omega = \frac{M_{i}^{2} - M_{f}^{2}}{2M_{i}}$, where $M_{i}$ and $M_{f}$ are the masses of the initial and final states. $\alpha$ is the fine structure constant, $\langle f\mid r\mid i\rangle$ is the transition matrix element and $C_{fi}$ is the statistical factor which is calculated using the 6J symbol, with $S=S_{i}=S_{f}$, as
\begin{equation*}
C_{fi} = \text{max}(L_{i}, L_{f})(2J_f+1)
        \begin{Bmatrix}
            L_{f} &  J_{f} & S\\
            J_{i} & L_{i} & 1 
        \end{Bmatrix}^{2}
\end{equation*}
In M1 transitions $|\Delta L| = 0$. Partial decay widths for M1 transitions are calculated using~\cite{PhysRevD.70.054017}\\
\begin{equation*}
    \Gamma_{M1}(i\rightarrow f\gamma) = \frac{ \alpha \mu^{2}\omega^{3}}{3}\left(2J_f+1\right)  \mid\langle f\mid j_{0}\left(\omega r/2\right)\mid i\rangle \mid^{2},
\end{equation*}
where $\alpha$ is the fine structure constant, $j_{0}(\omega r/2)$ is zeroth spherical Bessel function and $\mu = \frac{m_{\overline{b}}e_{c} - m_{c}e_{\overline{b}}}{4m_{\overline{b}}m_{c}}$ is the magnetic dipole moment. 
\\
Tables \ref{e1_1l} - \ref{m12} gives the $E1$ and $M1$ radiative decay widths along with the photon energies (in GeV) and absolute value of the transition matrix elements (in GeV$^{-1}$).

\begin{table}[h]
\caption{E1 radiative decay widths (in keV) of $n=1$, $L=1, 2$ states.\label{e1_1l}}
\begin{ruledtabular}
\begin{tabular}{cccccccccc}
$i$ & $f$ & $i \rightarrow f$ & $\omega$ & $\langle f|r| i\rangle$ & $\Gamma_{E1}$& \cite{QLI} & \cite{PhysRevD.70.054017} & \cite{ebert2003} & \cite{AKRAI}\\
\colrule
$1^3 P_0$ & $1^3 S_1$ & $1^3 P_0 \rightarrow 1^3 S_1$ & 0.340& -1.760& 72& 96& 55& 67.2& 58.55\\
$1^3 P_2$ & $1^3 S_1$ & $1^3 P_2 \rightarrow 1^3 S_1$ & 0.409& -1.760& 127& 87& 83& 107& 64.24\\
\multirow[t]{2}{*}{$1 P_1^{\prime}$} & $1^3 S_1$ & $1^3 P_1 \rightarrow 1^3 S_1$ & 0.403& -1.760& 57& 40& 11& 13.6& 9.98\\
& $1^1 S_0$ & $1^1 P_1 \rightarrow 1^1 S_0$ & 0.454& -1.605& 77& 74& 80& 132& 72.28\\
\multirow[t]{2}{*}{$1 P_1$} & $1^3 S_1$ & $1^3 P_1 \rightarrow 1^3 S_1$ & 0.389& -1.760& 58& 70& 60& 78.9& 51.14\\
& $1^1 S_0$ & $1^1 P_1 \rightarrow 1^1 S_0$ & 0.440& -1.605& 61& 35& 13& 18.4& 13.70\\
\multirow[t]{4}{*}{$1^3 D_1$} & $1^3 P_0$ & $1^3 D_1 \rightarrow 1{ }^3 P_0$ & 0.334& 2.779& 114& 65& 55& 128& 57.76\\
& $1^3 P_2$ & $1^3 D_1 \rightarrow 1^3 P_2$ & 0.264& 2.779& 2.8& 0.7& 1.8& 5.52& 2.15\\
& $1 P_1^{\prime}$ & $1^3 D_1 \rightarrow 1^3 P_1$ & 0.270& 2.779& 21& 12& 4.4& 7.66& 5.55\\
& $1 P_1$ & $1^3 D_1 \rightarrow 1^3 P_1$ & 0.285& 2.779& 28& 29& 28& 73.8& 29.61\\
$1^3 D_3$ & $1^3 P_2$ & $1^3 D_3 \rightarrow 1^3 P_2$ & 0.264& 2.779& 103& 67& 78& 102& 73.98\\
\multirow[t]{3}{*}{$1 D_2^{\prime}$} & $1^3 P_2$ & $1^3 D_2 \rightarrow 1^3 P_2$ & 0.274& 2.779& 19& 8.3& 8.8& 12.8& 20.57\\
& $1 P_1^{\prime}$ & $1^3 D_2 \rightarrow 1^3 P_1$ & 0.281& 2.779& 100& 41& 63& 116& \\
& $1 P_1$ & $1^3 D_2 \rightarrow 1^3 P_1$ & 0.295& 2.779& 2.2& 0.39& 7.0& 7.25& \\
\multirow[t]{3}{*}{$1 D_2$} & $1^3 P_2$ & $1^3 D_2 \rightarrow 1^3 P_2$ & 0.263& 2.779& 8.3& 8.7& 9.6& 27.5& 18.17\\
& $1 P_1^{\prime}$ & $1^1 D_2 \rightarrow 1^1 P_1$ & 0.269& -2.775& 7.0& 1.09& 15& 14.1& \\
& $1 P_1$ & $1^1 D_2 \rightarrow 1^1 P_1$ & 0.284& -2.775& 107& 44& 64& 112& \\
\end{tabular}
\end{ruledtabular}
\end{table}

\begin{table}[h]
\caption{E1 radiative decay widths (in keV) of $n=1$, $L=3$ states.\label{e1_1f}}
\begin{ruledtabular}
\begin{tabular}{cccccccc}
$i$ & $f$ & $i \rightarrow f$ & $\omega$ & $\langle f|r|i\rangle$ & $\Gamma_{E1}$ & \cite{QLI} & \cite{PhysRevD.70.054017} \\
\colrule
$1^3 F_2$ & $1^3 D_1$ & $1^3 F_2 \rightarrow 1^3 D_1$ & 0.245& 3.621& 125& 78& 75\\
 & $1^3 D_3$ & $1^3 F_2 \rightarrow 1^3 D_3$ & 0.244& 3.621& 0.65& 0.12& 0.4\\
 & $1 D_2^{\prime}$ & $1^3 F_2 \rightarrow 1^3 D_2$ & 0.234& 3.621& 13& 5.72& 6.3\\
 & $1 D_2$ & $1^3 F_2 \rightarrow 1^3 D_2$ & 0.246& 3.621& 7.7& 6.36& 6.5\\
$1{ }^3 F_4$ & $1{ }^3 D_3$ & $1^3 F_4 \rightarrow 1^3 D_3$ & 0.227& 3.621& 117& 69& 81\\
$1 F_3^{\prime}$ & $1^3 D_3$ & $1^3 F_3 \rightarrow 1^3 D_3$ & 0.248& 3.621& 10& 4.76& 3.7\\
 & $1 D_2^{\prime}$ & $1^3 F_3 \rightarrow 1^3 D_2$ & 0.238& 3.621& 126& 32& 78\\
 & $1 D_2$ & $1^3 F_3 \rightarrow 1^3 D_2$ & 0.250& 3.621& 1.5& 0.04& 0.5\\
$1 F_3$ & $1{ }^3 D_3$ & $1^3 F_3 \rightarrow 1^3 D_3$ & 0.226& 3.621& 5.2& 4.91& 5.4\\
 & $1 D_2^{\prime}$ & $1^1 F_3 \rightarrow 1^1 D_2$ & 0.217& -3.621& 0.19& 0.22& 0.04\\
 & $1 D_2$ & $1^1 F_3 \rightarrow 1^1 D_2$ & 0.229& -3.621& 114& 29& 82\\
\end{tabular}
\end{ruledtabular}
\end{table}

\begin{table}[h]
\caption{E1 radiative decay widths (in keV) of $n=2$, $L=0,1$ states.\label{e1_2l1}}
\begin{ruledtabular}
\begin{tabular}{cccccccccc}
$i$ & $f$ & $i \rightarrow f$ & $\omega$ & $\langle f|r| i\rangle$ & $\Gamma_{E1}$ & \cite{QLI} & \cite{PhysRevD.70.054017} & \cite{ebert2003} & \cite{AKRAI} \\
\colrule
\multirow[t]{4}{*}{$2{ }^3 S_1$} & $1^3 P_0$ & $2^3 S_1 \rightarrow 1{ }^3 P_0$ & 0.231& 2.166& 12& 3.48& 2.9& 3.78& 0.94\\
& $1^3 P_2$ & $2^3 S_1 \rightarrow 1{ }^3 P_2$ & 0.160& 2.166& 19& 6.98& 5.7& 5.18& 2.28\\
& $1 P_1^{\prime}$ & $2^3 S_1 \rightarrow 1{ }^3 P_1$ & 0.166& 2.166& 6.0& 1.56& 0.7& 0.63& 0.26\\
& $1 P_1$ & $2^3 S_1 \rightarrow 1^3 P_1$ & 0.181& 2.166& 8.8& 4.62& 4.7& 5.05& 1.45\\
\multirow[t]{2}{*}{$2{ }^1 S_0$} & $1 P_1^{\prime}$ & $2^1 S_0 \rightarrow 1^1 P_1$ & 0.134& 2.276& 12& 6.38& 6.1& 3.72& 3.03\\
& $1 P_1$ & $2{ }^1 S_0 \rightarrow 1{ }^1 P_1$ & 0.149& 2.276& 14& 5.33& 1.3& 1.02& 0.62\\
\multirow[t]{3}{*}{$2{ }^3 P_0$} & $2^3 S_1$ & $2^3 P_0 \rightarrow 2^3 S_1$ & 0.207& 2.792& 41& 53& 42& 29.2& 55.05\\
& $1^3 S_1$ & $2^3 P_0 \rightarrow 1^3 S_1$ & 0.750& 0.258& 17& 41& 1.0& & \\
& $1^3 D_1$ & $2^3 P_0 \rightarrow 1^3 D_1$ & 0.102& 2.333& 6.9& 5.6& 4.2& 0.036& 3.94\\
\multirow[t]{6}{*}{$2{ }^3 P_2$} & $2{ }^3 S_1$ & $2{ }^3 P_2 \rightarrow 2{ }^3 S_1$ & 0.274& 2.792& 96& 50& 55& 57.3& 64.92\\
& $1^3 S_1$ & $2^3 P_2 \rightarrow 1^3 S_1$ & 0.812& 0.258& 21& 52& 14& & \\
& $1^3 D_1$ & $2^3 P_2 \rightarrow 1^3 D_1$ & 0.170& 2.333& 0.3& 0.13& 0.1& 0.035& 0.07\\
& $1^3 D_3$ & $2^3 P_2 \rightarrow 1^3 D_3$ & 0.169& 2.333& 26& 14& 6.8& 1.59& 6.28\\
& $1 D_2^{\prime}$ & $2{ }^3 P_2 \rightarrow 1^3 D_2$ & 0.159& 2.333& 2.6& 0.93& 0.7& 0.113& 0.91\\
& $1 D_2$ & $2^3 P_2 \rightarrow 1^3 D_2$ & 0.172& 2.333& 1.6& 1.1& 0.6& 0.269& 1.16\\
\multirow[t]{7}{*}{$2 P_1^{\prime}$} & $2^3 S_1$ & $2^3 P_1 \rightarrow 2{ }^3 S_1$ & 0.268& 2.792& 40& 25& 5.5& 9.07& 15.11\\
& $1^3 S_1$ & $2^3 P_1 \rightarrow 1^3 S_1$ & 0.806& 0.258& 9.2& 26& 0.6& & \\
& $2{ }^1 S_0$ & $2{ }^1 P_1 \rightarrow 2{ }^1 S_0$ & 0.300& -2.539& 58& 36& 52& 72.5& 56.28\\
& $1^1 S_0$ & $2^1 P_1 \rightarrow 1^1 S_0$ & 0.854& -0.318& 21& 44& 19& & \\
& $1^3 D_1$ & $2^3 P_1 \rightarrow 1^3 D_1$ & 0.164& 2.333& 3.2& 1.27& 0.2& 0.073& 0.35\\
& $1 D_2^{\prime}$ & $2^3 P_1 \rightarrow 1^3 D_2$ & 0.153& 2.333& 19& 1.05& 5.5& 1.20& \\
& $1 D_2$ & $2^1 P_1 \rightarrow 1^1 D_2$ & 0.165& 2.338& 2.3& 0.03& 1.3& 0.149& \\
\multirow[t]{7}{*}{$2 P_1$} & $2^3 S_1$ & $2{ }^3 P_1 \rightarrow 2{ }^3 S_1$ & 0.252& 2.792& 41& 34& 45& 37.9& 50.40\\
& $1^3 S_1$ & $2{ }^3 P_1 \rightarrow 1^3 S_1$ & 0.791& 0.258& 11& 40& 5.4& & \\
& $2{ }^1 S_0$ & $2{ }^1 P_1 \rightarrow 2{ }^1 S_0$ & 0.283& -2.539& 39& 19& 5.7& 11.7& 16.52\\
& $1^1 S_0$ & $2^1 P_1 \rightarrow 1^1 S_0$ & 0.839& -0.318& 16& 25& 2.1& & \\
& $1^3 D_1$ & $2^3 P_1 \rightarrow 1^3 D_1$ & 0.148& 2.333& 3.0& 1.45& 1.6& 0.184& 1.14\\
& $1 D_2^{\prime}$ & $2^3 P_1 \rightarrow 1^3 D_2$ & 0.136& 2.333& 0.3& 0.006& 0.8& 0.021& \\
& $1 D_2$ & $2{ }^1 P_1 \rightarrow 1{ }^1 D_2$ & 0.149& 2.338& 18& 0.84& 3.6& 0.418& \\
\end{tabular}
\end{ruledtabular}
\end{table}

\begin{table}[h!]
\caption{E1 radiative decay widths (in keV) of $n=2$, $L=2$ states.\label{e1_2l2}}
\begin{ruledtabular}
\begin{tabular}{ccccccc}
$i$ & $f$ & $i \rightarrow f$ & $\omega$ & $\langle f|r| i\rangle$ & $\Gamma_{E1}$ &\cite{QLI} \\
\colrule
\multirow[t]{9}{*}{$2^3 D_1$} & $2{ }^3 P_0$ & $2^3 D_1 \rightarrow 2{ }^3 P_0$ & 0.271& -3.674& 107& 46\\
& $1^3 P_0$ & $2^3 D_1 \rightarrow 1^3 P_0$ & 0.686& 0.24& 7.4& 41.8\\
& $2^3 P_2$ & $2^3 D_1 \rightarrow 2{ }^3 P_2$ & 0.203& -3.674& 2.2& 0.58\\
& $1^3 P_2$ & $2^3 D_1 \rightarrow 1{ }^3 P_2$ & 0.619& 0.24& 0.3& 8.13\\
& $2 P_1^{\prime}$ & $2^3 D_1 \rightarrow 2{ }^3 P_1$ & 0.209& -3.674& 16& 10.15\\
& $1 P_1^{\prime}$ & $2^3 D_1 \rightarrow 1{ }^3 P_1$ & 0.625& 0.24& 2.0& 7.6\\
& $2 P_1$ & $2^3 D_1 \rightarrow 2^3 P_1$ & 0.226& -3.674& 26& 20.88\\
& $1 P_1$ & $2^3 D_1 \rightarrow 1{ }^3 P_1$ & 0.639& 0.24& 2.4& 12.5\\
& $1^3 F_2$ & $2^3 D_1 \rightarrow 1{ }^3 F_2$ & 0.128& -2.449& 3.8& \\
\multirow[t]{6}{*}{$2^3 D_3$} & $2{ }^3 P_2$ & $2{ }^3 D_3 \rightarrow 2{ }^3 P_2$ & 0.209& -3.674& 88& 54\\
& $1^3 P_2$ & $2^3 D_3 \rightarrow 1{ }^3 P_2$ & 0.624& 0.24& 10& 32\\
& $1^3 F_2$ & $2^3 D_3 \rightarrow 1{ }^3 F_2$ & 0.133& -2.449& 1.2& \\
& $1^3 F_4$ & $2^3 D_3 \rightarrow 1^3 F_4$ & 0.151& -2.449& 13& \\
& $1 F_3^{\prime}$ & $2^3 D_3 \rightarrow 1^3 F_3$ & 0.129& -2.449& 3.0& \\
& $1 F_3$ & $2^3 D_3 \rightarrow 1^3 F_3$ & 0.152& -2.449& 3.2& \\
\multirow[t]{8}{*}{$2 D_2^\prime$} & $2^3 P_2$ & $2^3 D_2 \rightarrow 2{ }^3 P_1$ & 0.214& -3.674& 16& 6.71\\
& $1^3 P_2$ & $2^3 D_2 \rightarrow 1{ }^3 P_1$ & 0.629& 0.24& 1.8& 7.28\\
& $2 P_1^{\prime}$ & $2{ }^3 D_2 \rightarrow 2{ }^3 P_1$ & 0.220& -3.674& 82& 29\\
& $2 P_1$ & $2^1 D_2 \rightarrow 2{ }^1 P_1$ & 0.237& -3.664& 3.7& 0.24\\
& $1 P_1^{\prime}$ & $2^3 D_2 \rightarrow 1{ }^3 P_1$ & 0.635& 0.24& 8.6& 19\\
& $1 P_1$ & $2^1 D_2 \rightarrow 1{ }^1 P_1$ & 0.649& -0.243& 0.2& 1.48\\
& $1 F_3^{\prime}$ & $2^3 D_2 \rightarrow 1^3 F_3$ & 0.135& 2.449& 8.9& \\
& $1 F_3$ & $2^1 D_2 \rightarrow 1^1 F_3$ & 0.157& -2.449& 0.1& \\
\multirow[t]{8}{*}{$2 D_2$} & $2^3 P_2$ & $2^3 D_2 \rightarrow 2^3 P_2$ & 0.206& -3.674& 6.5& 6.33\\
& $1^3 P_2$ & $2^3 D_2 \rightarrow 1{ }^3 P_2$ & 0.622& 0.24& 0.8& 7.04\\
& $2 P_1^{\prime}$ & $2^3 D_2 \rightarrow 2{ }^3 P_1$ & 0.212& -3.674& 8.3& 0.74\\
& $2 P_1$ & $2{ }^1 D_2 \rightarrow 2{ }^1 P_1$ & 0.229& -3.664& 96& 34\\
& $1 P_1^{\prime}$ & $2^3 D_2 \rightarrow 1^3 P_1$ & 0.628& 0.24& 0.8& 0.12\\
& $1 P_1$ & $2^1D_2 \rightarrow 1{ }^1 P_1$ & 0.642& -0.243& 9.3& 22.6\\
& $1 F_3^{\prime}$ & $2^3 D_2 \rightarrow 1{ }^3 F_3$ & 0.127& 2.449& 0.6& \\
& $1 F_3$ & $2^1 D_2 \rightarrow 1^1 F_3$ & 0.149& -2.449& 15& \\
\end{tabular}
\end{ruledtabular}
\end{table}

\begin{table}[h!]
\caption{E1 radiative decay widths (in keV) of $n=2$, $L=3$ states.\label{2f}}
\begin{ruledtabular}
\begin{tabular}{cccccc}
$i$ & $f$ & $i \rightarrow f$ & $\omega$& $\langle f|r| i\rangle$ & $\Gamma_{E1}$\\
\colrule
$2^3 F_2$ & $1^3 D_1$ & $2^3 F_2 \rightarrow 1^3 D_1$ & 0.245 & 3.621 & 125 \\
& $1^3 D_3$ & $2^3 F_2 \rightarrow 1^3 D_3$ & 0.244 & 3.621 & 0.6 \\
& $1 D_2^{\prime}$ & $2^3 F_2 \rightarrow 1^3 D_2$ & 0.234 & 3.621 & 14 \\
& $1 D_2$ & $2^3 F_2 \rightarrow 1^3 D_2$ & 0.246 & 3.621 & 7.7 \\
$2^3 F_4$ & $1^3 D_3$ & $2^3 F_4 \rightarrow 1^3 D_3$ & 0.227 & 3.621 & 117 \\
$2 F_3^{\prime}$ & $1^3 D_3$ & $2^3 F_3 \rightarrow 1^3 D_3$ & 0.248 & 3.621 & 10 \\
& $1 D_2^{\prime}$ & $2^3 F_3 \rightarrow 1^3 D_2$ & 0.238 & 3.621 & 126 \\
& $1 D_2$ & $2^3 F_3 \rightarrow 1^3 D_2$ & 0.25 & -3.621 & 1.5 \\
$2 F_3$ & $1^3 D_3$ & $2^3 F_3 \rightarrow 1^3 D_3$ & 0.226 & 3.621 & 5.2 \\
& $1 D_2^{\prime}$ & $2^1 F_3 \rightarrow 1^1 D_2$ & 0.217 & 3.621 & 0.2 \\
& $1 D_2$ & $2^1 F_3 \rightarrow 1^1 D_2$ & 0.229 & -3.621 & 115 \\
\end{tabular}
\end{ruledtabular}
\end{table}

\begin{table}[h!]
\caption{M1 radiative decay widths (in eV) of $n=1,2,3$, $L=0$ states.\label{m11}}
\begin{ruledtabular}
\begin{tabular}{ccccccccccc}
$i\rightarrow f$& $\omega$& $\langle i|\ldots|f \rangle $& $\Gamma_{M1}$& \cite{QLI}& \cite{PhysRevD.70.054017}& \cite{akbar_epja}& \cite{GONZ}& \cite{ebert2003}& \cite{ElHady2005radiativedecaybc}&\cite{soni}\\
\colrule
$1 ^3 S_1\rightarrow 1 ^1 S_0 \gamma$&0.054 & -0.383 & 57 & 57 & 80 & 27 & 52 & 33 & 18.9 & 53.109 \\
 $2 ^3 S_1\rightarrow 2 ^1 S_0 \gamma $&0.033 & -0.382 & 12 & 2.4 & 10 & 0.0016 & 10 & 17 & 3.7 & 21.119 \\
 $2 ^3 S_1\rightarrow 1 ^1 S_0 \gamma $&0.609 & -0.061 & 2042 & 1205 & 600 & 367 & 650 & 428 & 135.7 & 481.572 \\
 $2 ^1 S_0\rightarrow 1 ^3 S_1 \gamma $&0.529 & -0.008 & 64 & 99 & 300 & 6 & 250 & 488 & 163.8 & 568.346 \\
 $3 ^3 S_1\rightarrow 3 ^1 S_0 \gamma $&0.026 & 0.382 & 6.6 & 0.8 & 3 & 0.32 &   &   &   &   \\
 $3 ^3 S_1\rightarrow 2 ^1 S_0 \gamma $&0.436 & 0.072 & 1043 & 356 & 200 & 96 &   &   &   &   \\
 $3 ^3 S_1\rightarrow 1 ^1 S_0 \gamma $&0.980 & 0.041 & 3831 & 1885 & 600 & 431 &   &   &   &   \\
 $3 ^1 S_0\rightarrow 2 ^3 S_1 \gamma $&0.380 & -0.020 & 166 & 152 & 60 & 4.6 &   &   &   &   \\
 $3 ^1 S_0\rightarrow 1 ^3 S_1 \gamma $&0.910 & -0.008 & 400 & 510 & 4200 & 0.646 &   &   &   &   \\
\end{tabular}
\end{ruledtabular}
\end{table}

\begin{table}[h!]
\caption{M1 radiative decay widths (in eV) of $n=4,5$, $L=0$ states.\label{m12}}
\begin{ruledtabular}
\begin{tabular}{cccccccccc}
$i\rightarrow f$&  $\omega$&  $\langle i|\ldots|f \rangle $&  $\Gamma_{E1}$& \cite{QLI} & $i\rightarrow f$& $\omega$& $\langle i|\ldots|f \rangle $& $\Gamma_{M1}$&\cite{QLI} \\
\colrule
$4 ^3 S_1\rightarrow 4 ^1 S_0 \gamma$&  0.023&  -0.381&  4.5&  0.35& $5 ^1 S_0\rightarrow 2 ^3 S_1 \gamma $& 0.990& -0.017& 2116& 1260\\
 $4 ^3 S_1\rightarrow 3 ^1 S_0 \gamma$&  0.369&  -0.081&  805&  252& $5 ^1 S_0\rightarrow 1 ^3 S_1 \gamma $& 1.475& -0.009& 1967& 1893\\
 $4 ^3 S_1\rightarrow 2 ^1 S_0 \gamma$&  0.760&  -0.049&  2548&  806& $6 ^3 S_1\rightarrow 6 ^1 S_0 \gamma$& 0.020& -0.381& 2.8&0.18\\ 
 $4 ^3 S_1\rightarrow 1 ^1 S_0 \gamma$&  1.279&  -0.033&  5644&  2501& $6 ^3 S_1\rightarrow 5 ^1 S_0 \gamma$& 0.303& -0.095& 611&191\\ 
 $4 ^1 S_0\rightarrow 3 ^3 S_1 \gamma$& 0.321& 0.030& 225& 186& $6 ^3 S_1\rightarrow 4 ^1 S_0 \gamma$& 0.600& -0.063& 2100&643\\
 $4 ^1 S_0\rightarrow 2 ^3 S_1 \gamma$& 0.709& -0.018& 833& 579& $6 ^3 S_1\rightarrow 3 ^1 S_0 \gamma$& 0.921& -0.047& 4181&1239\\
 $4 ^1 S_0\rightarrow 1 ^3 S_1 \gamma$& 1.214& -0.009& 1023& 1122& $6 ^3 S_1\rightarrow 2 ^1 S_0 \gamma$& 1.284& -0.035& 6490&1917\\
 $5 ^3 S_1\rightarrow 5 ^1 S_0 \gamma$& 0.021& -0.381& 3.4& 0.18& $6 ^3 S_1\rightarrow 1 ^1 S_0 \gamma$& 1.765& -0.027& 9625&3772\\
 $5 ^3 S_1\rightarrow 4 ^1 S_0 \gamma$& 0.329& -0.089& 687& 210& $6 ^1 S_0\rightarrow 5 ^3 S_1 \gamma$& 0.263& -0.047& 291&225\\
 $5 ^3 S_1\rightarrow 3 ^1 S_0 \gamma$& 0.661& -0.057& 2257& 675& $6 ^1 S_0\rightarrow 4 ^3 S_1 \gamma$& 0.560& -0.034& 1488&849\\
 $5 ^3 S_1\rightarrow 2 ^1 S_0 \gamma$& 1.037& -0.040& 4372& 1316& $6 ^1 S_0\rightarrow 3 ^3 S_1 \gamma$& 0.879& 0.025& 3141&1613\\
 $5 ^3 S_1\rightarrow 1 ^1 S_0 \gamma$& 1.536& -0.029& 7564& 3107& $6 ^1 S_0\rightarrow 2 ^3 S_1 \gamma$& 1.239& -0.017& 4127&2203\\
 $5 ^1 S_0\rightarrow 4 ^3 S_1 \gamma $& 0.286& -0.039& 264& 209& $6 ^1 S_0\rightarrow 1 ^3 S_1 \gamma$& 1.707& -0.009& 3272&2822\\
 $5 ^1 S_0\rightarrow 3 ^3 S_1 \gamma $& 0.617& 0.026& 1194& 720& & & & &\\
\end{tabular}
\end{ruledtabular}
\end{table}

\section{Discussion and Conclusion}\label{sec 4}
In this work we have used the matrix Numerov method to solve the Schrodinger equation for $B_c$ mesons. The model potential consists of the familiar Cornell potential along with an $\mathcal{O}(1/m)$ correction from LQCD. The hyperfine interaction term has been considered non-perturbatively in our analysis. We have solved the matrix eigenvalue problem and computed the spectra of $S$, $P$, $D$ and $F$ wave states and their corresponding wave functions. The masses are presented in Tables \ref{tab_sp} and \ref{tab_df} in comparison with some relativistic \cite{PhysRevD.70.054017, XJLi} and nonrelativistic \cite{QLI, soni} potential models. In Ref.\cite{ABREU}, $Bc$ spectra has been obtained using the Tamm-Dancoff approximation. The masses obtained in the present work are comparable with the predictions from other theoretical studies. Also our masses for $Bc(1^3S_1)$, $ Bc(1^3P_0)$ and $ Bc(1P)$ states are good in agreement with the lattice results \cite{Lat}. From our results, we obtain the mass difference $B_c^{*+} (2S)-B_c^{*+} (1S)$ to be 584 MeV which is close to the experimental difference of 567 MeV \cite{disc2}. Presently, a thorough analysis of the spectra cannot be done due to lack of experimental data. 

Table \ref{tab_rms} gives the rms radii of $Bc$ mesons for various radial and orbital excitations. We see that vector mesons have slightly higher rms radii than the corresponding pseudoscalar mesons. Furthermore, the rms radii increases with increase in radial and orbital quantum numbers. Charmonium \cite{bhagAHEP, rmscc} and bottomonium \cite{rmsbb1, manzoor2021newvariational} systems also exhibit similar trend in rms radii.  Our results for decay constants are consistent with results from nonrelativistic models \cite{akb2,soni,2} and other theoretical approaches like light front formalism \cite{3} and QCD sum rules \cite{4,5}. From Table \ref{dec_con} we see that $f_v < f_p$ and the decay constants decreases with increase in radial quantum number. Thus, we see that while the rms radii increases and decay constants decreases as the radial quantum number increases. This trend as suggested in Ref.\cite{3} may be due to the fact that as radial quantum number increases, the mesons are loosely bound and hence a larger spread in radial distribution. Similar trend is also seen in charmonium and bottomonium systems.

We have calculated weak decay widths and lifetimes following the spectator model. The weak decay widths, branching ratios and lifetimes for $1^1S_0$ state are listed in Table~\ref{weak}. We see that the branching ratio for $b$ quark decay is greater than that for $c$ quark decay and annihilation decay. This result is in agreement with the predictions from other models. Knowing the lifetime of the particle helps us understand the weak interaction properties. The obtained lifetime of the pseudoscalar $1^1S_0$ state is $0.41$ ps which is comparable with the experimental lifetime of $0.510\pm 0.009$ ps \cite{pdg}.

The decay widths for radiative E1 transitions for ground and various radially and orbitally excited states are given in Tables \ref{e1_1l}-\ref{2f} along with the photon energies and the overlap integrals. The decay widths are also compared with the predictions from other models and our results are in reasonable agreement with these models. The M1 radiative widths of lower and higher $S$ wave states are given in Tables \ref{m11} and \ref{m12} along with the photon energies and the overlap integrals. We have compared the widths with other relativistic and nonrelativistic models. In the nonrelativistic limit, the transitions for which the radial quantum number ($n$) changes are hindered because of the wavefunction orthogonality. However, from our results, we see that these hindered transitions have widths comparable to the allowed transitions. In our model, we have treated the spin-spin interaction non-perturbatively, and hence the singlet and triplet states for the same $n$ will have slightly different wavefunctions. As a result, the wavefunction orthogonality will be broken as suggested in Ref.\cite{PhysRevD.70.054017}. Another reason for the appreciable widths of the hindered transitions is the large $\omega^3$ dependence of the M1 decay width. 
\\

To sum up, the current article applied the matrix Numerov method to study the spectra and decay properties of $B_c$ mesons. The findings of this study are consistent with both available experimental data and predictions from other theoretical frameworks.
\def\bibsection{\section*{\refname}} 
\bibliography{ref}

\end{document}